\documentclass[usenatbib]{mn2e}
\usepackage{graphicx,fixltx2e}

\def\msun{\hbox{${\rm M}_{\odot}$}}
\def\mspy{\hbox{${\rm M}_{\odot}$\,yr$^{-1}$}}
\def\rsun{\hbox{${\rm R}_{\odot}$}}

\def\mstar{\hbox{$M_{\star}$}}
\def\rstar{\hbox{$R_{\star}$}}

\def\eps{\hbox{erg\,s$^{-1}$}}

\def\pcc{\hbox{cm$^{-3}$}}

\def\em{\it}

\def\degr{\hbox{$^\circ$}}

\def\Mdot{\hbox{$\dot{M}$}}

\newcommand{\bvec}[1]{\mbox{\boldmath ${#1}$}}
\newcommand{\deriv}[2]{\mbox{${{\displaystyle d#1}\over {\displaystyle d#2}}$}}

\newcommand{\caii}{\hbox{Ca$\;${\sc ii}}}

\begin{document}

\title[Coronal structure of the cTTS V2129~Oph]{Coronal structure of the classical T~Tauri star V2129~Oph\thanks{Based on observations 
obtained at the Canada-France-Hawaii Telescope (CFHT) which is operated by the National 
Research Council of Canada, the Institut National des Sciences de l'Univers of the Centre 
National de la Recherche Scientifique of France, and the University of Hawaii.} }

\makeatletter

\def\newauthor{%
  \end{author@tabular}\par
  \begin{author@tabular}[t]{@{}l@{}}}
\makeatother
 
\author[M.M.~Jardine et al.]
{\vspace{1.7mm}
M.M.~Jardine$^1$\thanks{E-mail: mmj@st-andrews.ac.uk (MMJ);
sg64@st-andrews.ac.uk (SGG); 
donati@ast.obs-mip.fr (J-FD)}, 
S.G.~Gregory$^1$, J.-F.~Donati$^2$ \\ 
\vspace{1.7mm}
{\hspace{-1.5mm}\LARGE\rm 
 } \\ 
$^1$ School of Physics and Astronomy, Univ.\ of St~Andrews, St~Andrews, Scotland KY16 9SS, UK \\
$^2$ LATT, CNRS--UMR 5572, Obs.\ Midi-Pyr\'en\'ees, 14 Av.\ E.~Belin, F--31400 Toulouse, France
}

\date{2007, MNRAS, submitted}
\maketitle
 
\begin{abstract} 
The nature of the magnetic coupling between T Tauri stars and their disks determines not only the mass accretion process but possibly the spin evolution of the central star. We have taken a recently-published surface magnetogram of one moderately-accreting T Tauri star (V2129 Oph) and used it to extrapolate the geometry of its large-scale field. We determine the structure of the open (wind-bearing) field lines, the closed (X-ray bright) field lines and the relatively small subset of field lines that pass through the equatorial plane inside the Keplerian co-rotation radius and which are therefore available to accrete. 

We consider a series of models in which the stellar magnetic field is opened up by the outward pressure of the hot coronal gas at a range of radii or {\em source surfaces}. As the source surface is increased, accretion takes place along progressively simpler field structures and impacts on progressively fewer sites at the stellar surface. This is consistent with the observed variation in the Ca II IRT and HeI lines which suggests that accretion in the visible hemisphere is confined to a single high-latitude spot. By determining the density and velocity of the accretion flows, we find that in order to have most of the total mass accretion rate impacting on a single high-latitude region we need disk material to accrete from approximately 7R$_\star$, close to the Keplerian co-rotation radius at 6.8R$_\star$. 

We also calculate the coronal density and X-ray emission measure. We find that both the magnitude and rotational modulation of the emission measure increase as the source surface is increased. For the field structure of V2129 Oph which is dominantly octupolar, the emission forms a bright, high-latitude ring that is always in view as the star rotates. Since the accretion funnels are not dense enough to cause significant scattering of coronal X-ray photons, they provide only a low rotational modulation of around 10$\%$ at most.

\end{abstract}


\section{Introduction} 
T Tauri stars are young solar-like stars that are still in the process of contracting onto the main sequence. The so-called classical T Tauri  stars (cTTS) show signs of active accretion and are typically younger than their counterparts the weak-line T Tauri stars. One of the enduring puzzles about cTTS is their slow to moderate rotation, which suggests that they must have succeeded in losing angular momentum as they contracted. This can in principle be achieved by magnetically coupling the star to its disk and indeed the presence of a disk, as indicated by IR excesses does seem to be correlated with slower rotation (e.g. \citet{rebull_onc_2006}). While the detailed mechanisms by which this coupling is achieved are still unclear, the general hypothesis is well established. The stellar magnetic field  disrupts the inner edge of the disk, forcing accreting material to flow along field lines down to the stellar surface. The torque associated with this magnetic field may balance the accretion torque, thus preventing the star from spinning up. Any attempt by the star to spin up generates a stronger magnetic field which exerts a  greater torque that acts to slow the star down. Field strengths at the stellar surface can then be estimated by calculating the values at the disk necessary for such a torque balance and extrapolating back to the surface, assuming some simple geometry  \citep{konigl91,cameron_campbell_93,shu94}. This process, however, requires surface field strengths of around 0.1 - 5 kG and a disk that is truncated close to the Keplerian corotation radius. 

The models can in principle be tested by measuring the surface field strength of T Tauri stars.  Zeeman-Broadening measurements can now be used to measure mean fields on the stellar surface \citep{johns_krull_NTT_04,valenti_johns_krull_04,yang_johns_krull_TWHya_05,johns_krull_07}.  This gives distributions of local field strengths on the stellar surface that may be as high as 6kG which is well within the range predicted by theory. However, these are similar on both accreting and non-accreting T Tauri stars, suggesting that disk-locking is not responsible for determining the field strength. There also seems to be little correlation with either the values predicted by theory, or with the stellar rotation rotation rate or Rossby number. Furthermore, although the average surface field strengths reach several kG, current 
observations indicate that the strength of the dipole component is at least an order of magnitude below what 
is required for disc-locking models to operate (e.g. \citealt*{daou_TT_06}; \citealt*{yang_TWHya_07}).

%% fig1 
\begin{figure}
%\center{\includegraphics[width=7.5cm]{meanB_rad_nonrad.pdf}} 
\center{\includegraphics[width=7.5cm]{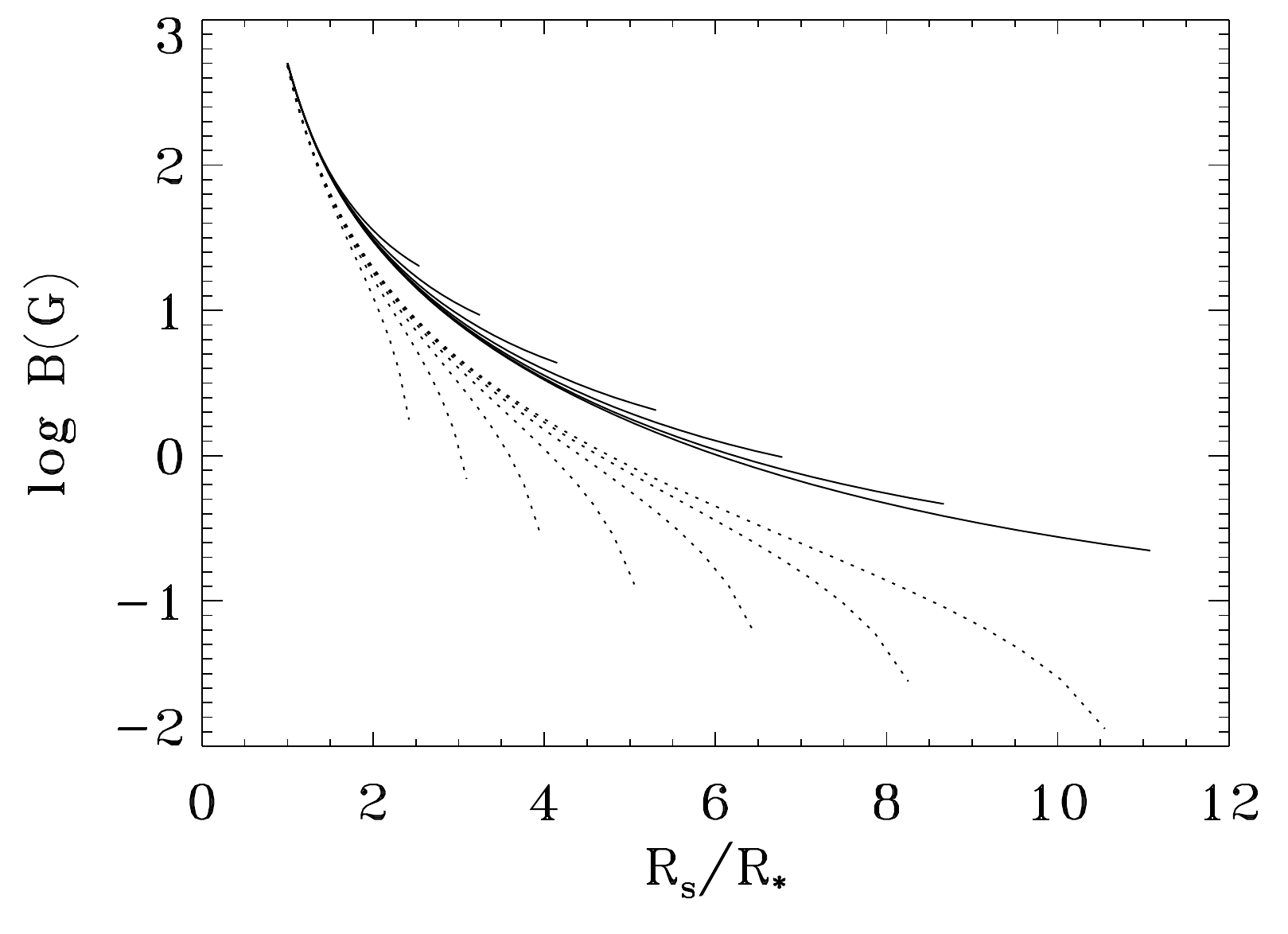}} 
\caption[]{The variation with radius of (solid) the radial component of the magnetic field and (dotted) the non-radial component. In both cases, the value plotted is averaged over latitude and longitude. Lines are drawn for a range of source surfaces.  At the source surface itself, both $B_\theta$ and $B_\phi$ are identically zero.}
\label{MeanB}
\end{figure}

A complementary approach is to use circular polarisation measurements both in photospheric lines (presumably formed across the whole stellar surface) and the Helium $\lambda$5876{\AA} emission line believed to form in the accretion shock  \citep{beristain_01,yang_TWHya_07}.  Polarisation measurements taken at a range of rotation phases can then be used to provide information on the structure of the magnetic field both across the surface and in the accretion shock. The low circular polarisation measured in photospheric lines rules out the presence of a global dipole, suggesting perhaps a complex field that may be locally very intense, but which is organised into small enough scales that the Zeeman signatures are complex in shape and associated with small longitudinal fields. In comparison, the large, rotationally modulated unipolar Zeeman signatures in the Helium emission lines suggest that the field associated with accretion is much simpler. This is no doubt to be expected, since it is the largest-scale field lines that will interact with the disk and these are likely to have the simplest structure. The nature of the star-disk interaction will clearly depend on the structure of the whole coronal field, and in particular the degree to which the complexity that seems apparent at the stellar surface may have died away by the time the coronal field reaches the inner edge of the disk.
 
  %% fig2 
\begin{figure}
%\center{\includegraphics[width=7.cm]{m_3025all_lon32.pdf}} 
%\center{\includegraphics[width=7.cm]{m_3040all_lon32.pdf}} 
\center{\includegraphics[width=7.cm]{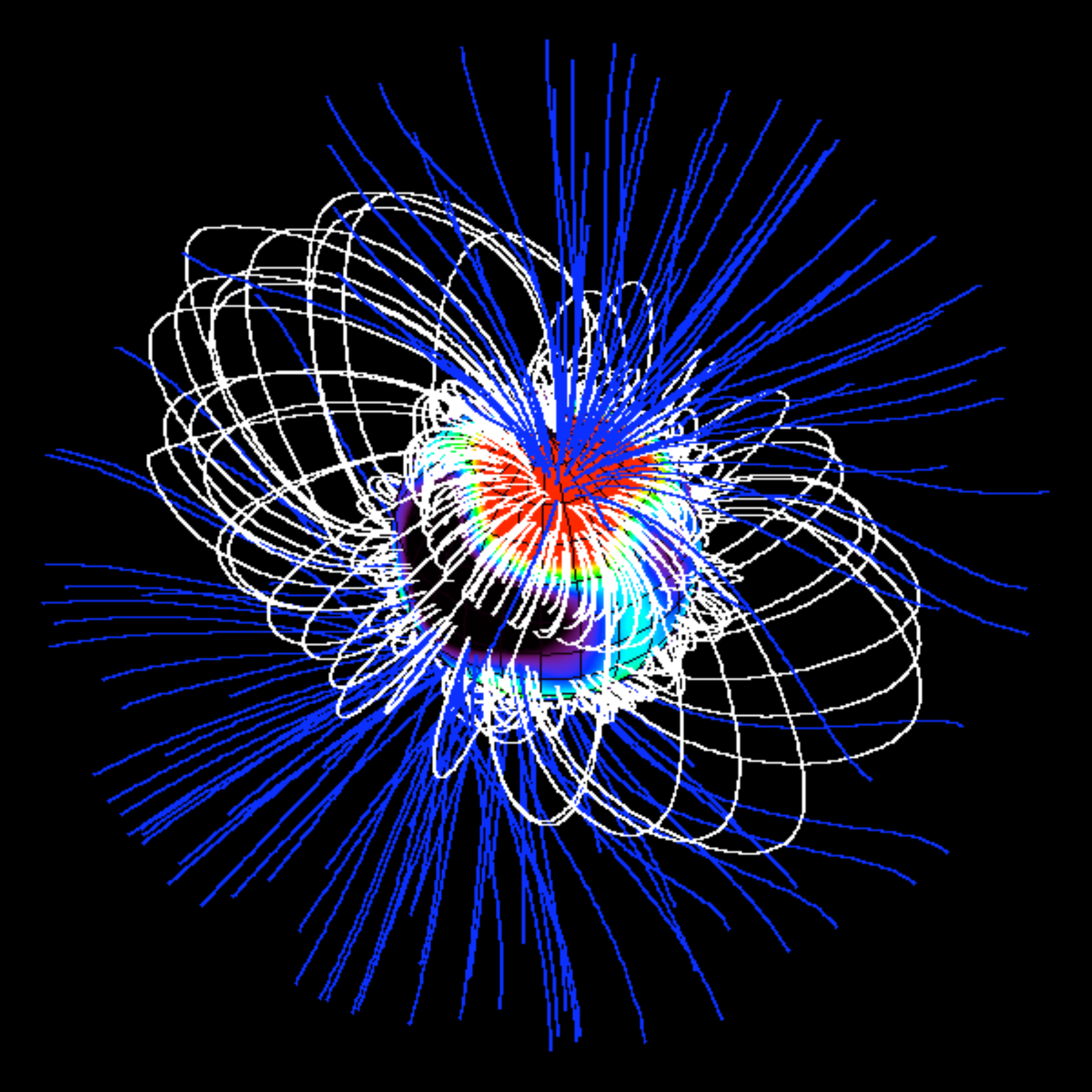}} 
\center{\includegraphics[width=7.cm]{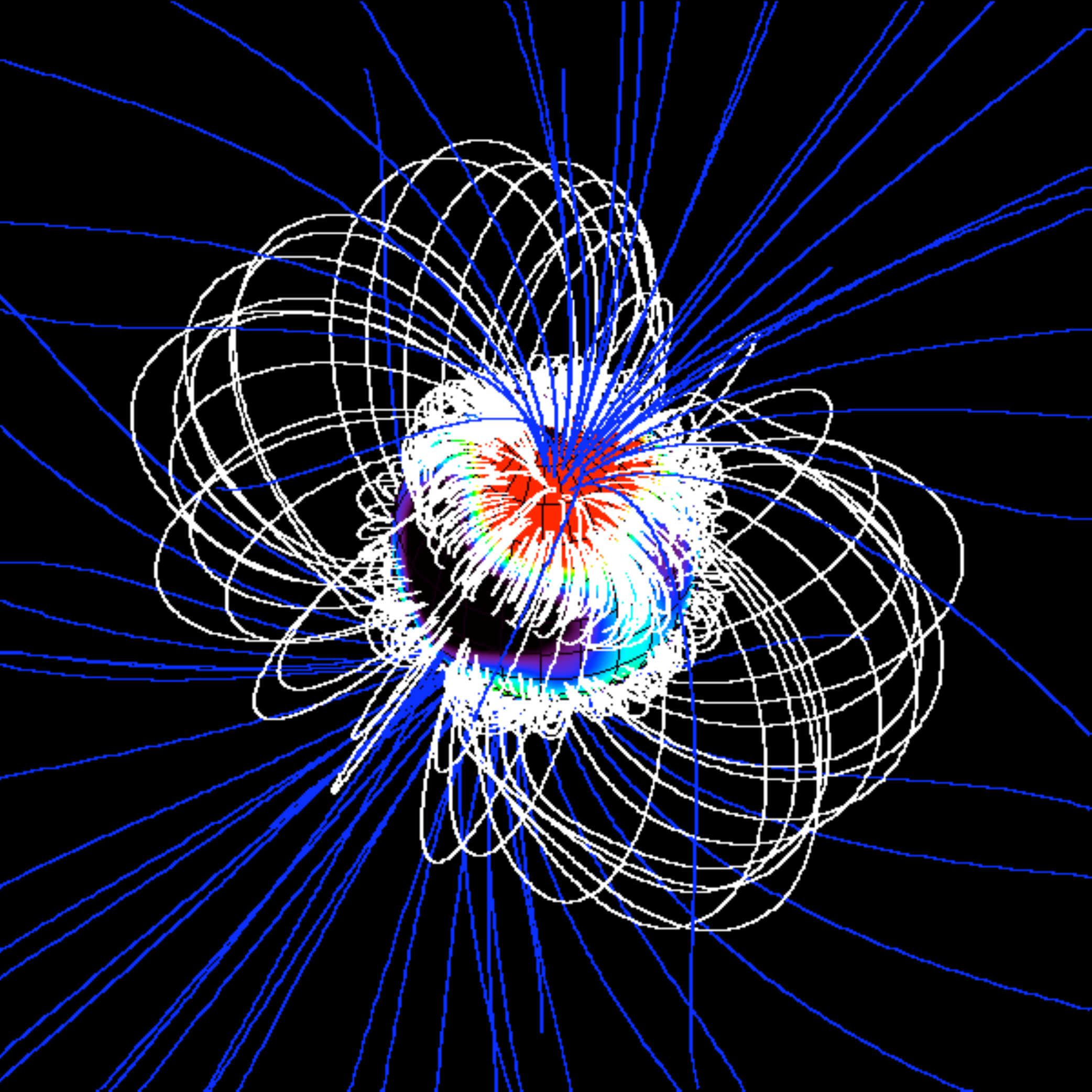}} 
\caption[]{The coronal magnetic structure of V2129 Oph for a source surface at 3.2R$_\star$ (top) and at 6.8R$_\star$ (bottom). Closed field lines are shown white, open field lines are shown blue. }
\label{fieldlines}
\end{figure}

 Signatures of this complexity may be apparent in the X-ray emission from T Tauri stars. In addition to the ubiquitous flaring-like variability seen in the X-ray emission of the {\it Chandra} Orion Ultradeep 
Project (COUP; Getman et al. 2005
%\citealt{getman_COUP_list_05a}
) stars, there is also a clear rotational modulation in a small but significant number of these stars \citep{flaccomio_COUP_rotmod_05}. Such modulation has previously been difficult to detect because it requires continuous observation over several rotation periods \citep{gudel95,marino03,hussain_chandra1_05}. This rotational modulation suggests that the X-ray coronae of T Tauri stars are compact, with discrete bright and dark regions. Some contribution to the rotational modulation may also come from the accretion funnels themselves, which, if they are dense enough, may obscure the underlying emission. \citet{gregory_rotmod_06} have modelled the structure of T  Tauri coronae and found significant rotational modulation, with X-ray periods that are typically equal to the optical period (or one half of it). While there seems to be little time-correlation between optical and X-ray variability, there appears to be a significant correlation between optical variability and X-ray luminosity. This suggests that  the X-ray emission of T Tauri stars is primarily produced in a complex arrangement of magnetically-active regions in a manner similar to observed on the Sun \citep{stassun_timevar_06,stassun_xray_opt_var_07}.
  
One singularly non-solar aspect of T Tauri coronae however is the possible interaction between the magnetic field of the star and the accretion disk.  Both in the COUP study and also more recently in the DROXO (Deep Rho Ophiuchi Xmm-Newton Observation) study, there are flares for which modelling suggests the loops responsible must be large - perhaps reaching out into the accretion disk \citep{favata_COUP_flares_05}. Field lines that connect the star and the disk in this way will be stressed by their interaction with the disk and may indeed carry much of the accretion torque \citep{rekowski_star_disc_04,long_nondipole_07}. The nature of the star-disk interaction may therefore depend on the structure of the magnetic field that links to the disk, whether it is primarily open as in these more recent models, or closed as in the traditional disk-locking models  \citep{konigl91,cameron_campbell_93,shu94,romanova_02,romanova_03,long_disk_locking_05}. Recent models of accretion powered stellar winds suggest that the angular momentum gained by the star due to accretion may be lost in a stellar wind, rather than exported back into the disk \citep{matt_pudritz_torques_08,matt_pudritz_eqm_08}. The degree of complexity of the field affects the torque that it can exert and the location of the inner edge of the accretion disk. 

Recent results by \citet{jardine_tts_06} suggest that the nature of this star-disk interaction may be fundamentally different in high and low mass stars. If we consider pre-main sequence stars of progressively lower mass, their  pressure scale heights $\Lambda$ will {\it increase} (since for a polytrope $R \propto M^{-1/3}$ and so $\Lambda \propto M_*^{-5/3}$) while their co-rotation radii will {\it decrease}. Thus the lower the stellar mass, the more likely it is that the co-rotation radius will be within the corona. If a disk is present, this may mean that the disk strips off the outer edge of the corona,  converting field lines that were closed and therefore possibly X-ray bright, to an open, X-ray dark, wind-bearing configuration. Even in the absence of a disk, of course, centrifugal effects may do the same job, preventing the corona from extending much beyond the co-rotation radius \citep{safier_98}. This may well explain the drop in X-ray emission measures towards lower masses seen in the COUP dataset (Getman et al. 2005)
%\citep{getman_COUP_list_05a}, 
but it also suggests that for the higher-mass stars, whose coronae may not extend as far as the co-rotation radius, any disk that exists at the co-rotation radius may interact not with the closed X-ray emitting field lines of the star but rather with the open field, which in the absence of a disc would carry a stellar wind. It is not clear what effect this would have on the interaction of the stellar and disk magnetic fields, and in particular on the location of the inner edge of the accretion disk. This may be determined by a balance between the matter energy density of the disk and the magnetic energy density of the stellar magnetic field, or by a small-scale balance between the disk viscosity and the resistivity which allows matter to transfer from disk field lines to stellar field lines (e.g. \citet{romanova_02,  cameron_campbell_93}). It seems clear that not only the strength of the stellar field but also its structure are important. Recent models by \citet{bessolaz_08} suggest that accretion funnels may indeed form in mildly-accreting T Tauri stars with weak dipolar fields of only 140G, but that higher field strengths or different field topologies are needed to explain more typical accretion rates of $\Mdot\simeq10^{-8}$~\mspy .  

In this paper we use recently-published surface magnetograms of a fairly typical T Tauri star to determine the structure of its corona and in particular to investigate the relationship between the closed-field regions that may contribute to the X-ray emission, the open field lines that may contribute to the wind and the field lines that pass through the equatorial plane inside of co-rotation which are available to accrete. By solving for the flow along these accreting field lines we determine the influence of the accretion funnels on the predicted magnitude and rotational modulation of the X-ray emission and also the mass accretion rate.

\section{The cTTS V2129~Oph}

V2129~Oph (SR~9, AS~207A, GY~319, ROX~29, HBC~264) is a fairly typical, moderately accreting classical T Tauri star with rotation rate of 6.53 d \citep{shevchenko_98}. Many of its parameters are derived in \citet{donati_v2129oph_07} and we simply summarise their results here. They find a radius and mass of $\rstar=2.4\pm0.3$~\rsun and $\mstar=1.35\pm0.15$~\msun\  (which imply that the co-rotation radius is at 6.8$\rstar$) and using the evolutionary models of \citet{siess_evol_tracks_00}, they obtain an age of about 2~Myr and conclude that V2129 Oph is no longer fully convective, with a small radiative core of mass $\simeq0.1$~\mstar\ and radius $\simeq0.2$~\rstar.  The presence of a radiative core is relevant to the nature of magnetic field generation since it provides a region of high shear at the boundary between the core and the convective region within which field can be generated. The mass accretion rate may be derived from the equivalent width of the 8662{\AA} \caii\ line emission core, giving $4\times10^{-9}$~\mspy\ \citep{mohanty_lowmass_05}.  \citet{eisner_05} suggests that \Mdot\ is significantly larger, about $3\times10^{-8}$~\mspy.  We follow \citet{donati_v2129oph_07} and assume here that $\Mdot\simeq10^{-8}$~\mspy, i.e., a conservative compromise between both estimates and a typical value for a mildly accreting cTTSs.  

The surface magnetic field maps derived by  \citet{donati_v2129oph_07} are based on circular polarisation Zeeman signatures detected in both photospheric lines and also in emission lines tracing magnetospheric accretion. The surface field that is recovered is complex, with many mixed-polarity regions. It is dominated by a 1.2kG octupole, tilted by about 20$^\circ$ to the rotation axis. A weaker dipole component is also present, with polar strength of 0.35kG and a similar tilt of about 30$^\circ$. For the hemisphere in view, accretion appears to be concentrated in a single high-latitude region which covers about 5$\%$ of the stellar surface and is in view at around rotation phase 0.8. This apparently coincides with a dark polar feature at the photospheric level. Observational evidence that the visible accretion spot is located at high 
latitudes is strong;  the low amplitude fluctuations in the radial 
velocity of the He I line (or the Ca II IRT emission core) on the one 
hand, and the fact that the accretion spot (and associated Zeeman 
signature) is always visible to the observer on the other hand, both argue 
in favour of this interpretation. 
This implies that any accretion from the disk must flow up out of the equatorial plane in order to impact the stellar surface at high latitudes. 

\section{Extrapolating from the surface field}
Once the surface magnetic field is known it is possible to extrapolate the coronal magnetic field using a variety of methods and assumptions. Two of the most commonly-used are a full MHD solution and the ``Potential Field Source Surface'' method which was originally developed by \citet{altschuler69} for extrapolating the Sun's coronal field from solar magnetograms. The latter has the advantage of computational speed and simplicity, but it cannot reproduce time-dependent or non-potential effects, both of which are likely to be important for the structure of the field in the region of the accretion disk. The former method can treat both of these effects, but is computationally very expensive and requires some assumption of an equation of state (often a polytrope is used) and the nature of the energetics. A review and comparison of the two techniques is given in \citet{riley_PFSS_06}.

While a full MHD solution is desirable in order to determine the nature of the interaction of the stellar field with the accretion disk, we postpone this approach for the moment, and concentrate on the first question that should be asked: is this surface magnetogram consistent with the inferred location of the accretion flows? These flows originate in the accretion disk and if they are magnetically-channelled, the location at which they arrive at the stellar surface must  depend on the structure of the coronal magnetic field. Only some fraction of the field lines that leave the stellar surface ever come close enough to the accretion disk to interact with it. This interaction is likely to be time-dependent and to inject stresses into the magnetic field, just as the convective motions, surface differential rotation and meridional flows stress the magnetic field lines that emerge through the stellar surface. Nonetheless, the subset of stellar field lines that pass through the disk and so are available to accrete is likely to be determined principally by the nature of the stellar field. We concentrate therefore on determining this overall structure and on determining which field lines are available for accretion and where they connect to the stellar surface. 

%% fig3 
\begin{figure}
%\center{\includegraphics[width=7.5cm]{B_disk.pdf}} 
\center{\includegraphics[width=7.5cm]{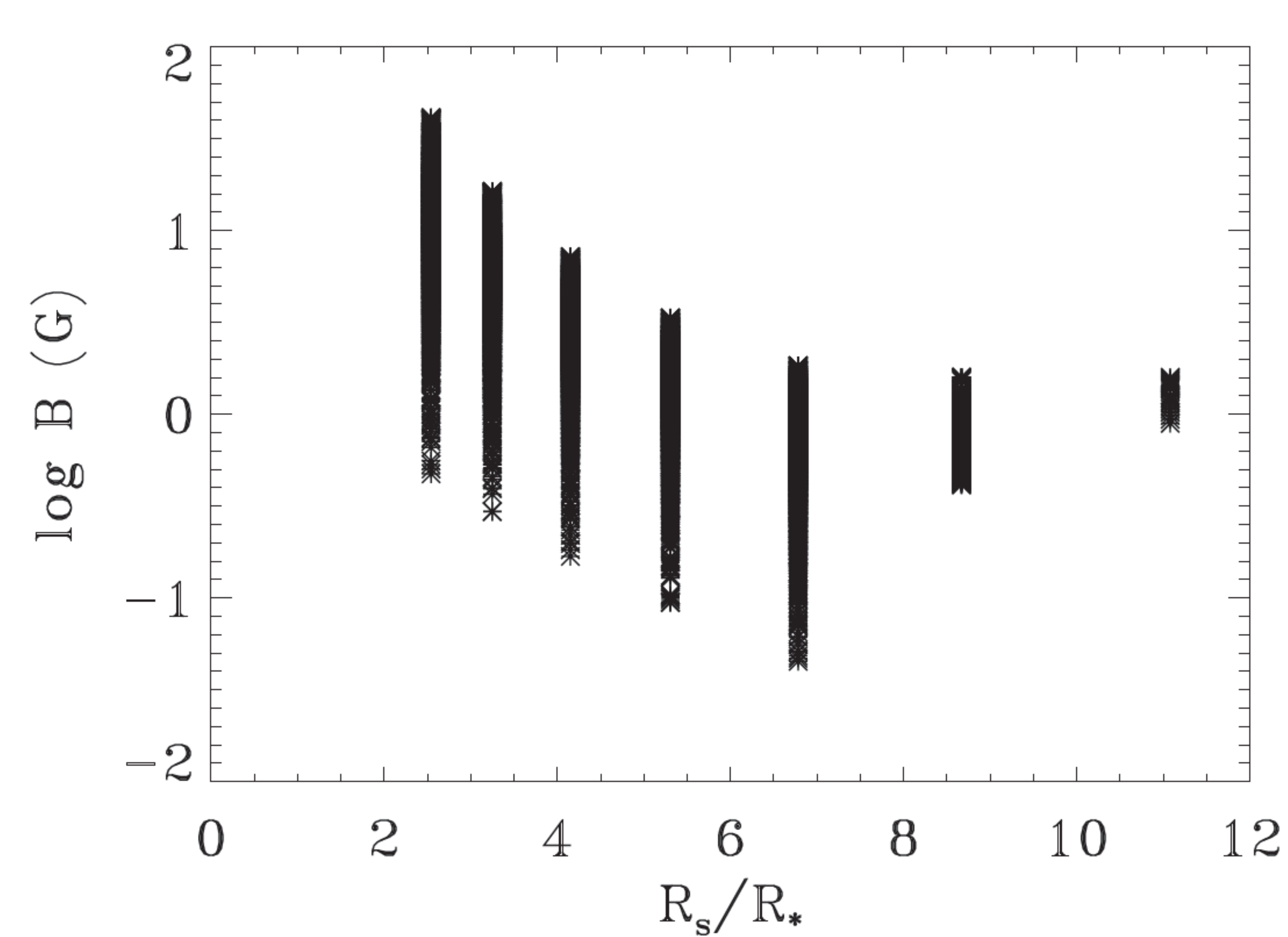}} 
\caption[]{Magnetic field strength of a range of field lines at the point where disk material latches onto the field. If the source surface is beyond the co-rotation radius, this location is taken to be the co-rotation radius. if, on the other hand, the source surface is inside co-rotation, this location is taken to be the source surface itself. Results are shown for a range of source surface locations.  }
\label{B_disk_nopress}
\end{figure}

We therefore use the ``Potential Field Source Surface'' method with a code originally developed by \citet{vanballegooijen98}. Since the method has been described in \citet{jardine02structure} we provide only an outline here. Briefly, we
write the magnetic field $\bvec{B}$ in terms of a flux function $\Psi$
such that $\bvec{B} = -\bvec{\nabla} \Psi$ and the condition that the
field is potential ($\bvec{\nabla}\times\bvec{B} =0$) is satisfied
automatically.  The condition that the field is divergence-free then
reduces to Laplace's equation $\bvec{\nabla}^2 \Psi=0$ with solution 
in spherical co-ordinates $(r,\theta,\phi)$
\begin{equation}
 \Psi = \sum_{l=1}^{N}\sum_{m=-l}^{l} [a_{lm}r^l + b_{lm}r^{-(l+1)}]
         P_{lm}(\theta) e^{i m \phi},
 \label{psi}
\end{equation}
where the associated Legendre functions are denoted by $P_{lm}$.  
This then gives
%\begin{equation}
$$
B_r  =  -\sum_{l=1}^{N}\sum_{m=-l}^{l}
               [la_{lm}r^{l-1} - (l+1)b_{lm}r^{-(l+2)}]
               P_{lm}(\theta) e^{i m \phi}
$$
%\end{equation}
%\begin{equation}
$$
B_\theta  =  -\sum_{l=1}^{N}\sum_{m=-l}^{l} 
               [a_{lm}r^{l-1} + b_{lm}r^{-(l+2)}]
               \deriv{}{\theta}P_{lm}(\theta) e^{i m \phi}
$$
%\end{equation} 
%\begin{equation} 
$$
B_\phi  =  -\sum_{l=1}^{N}\sum_{m=-l}^{l} 
               [a_{lm}r^{l-1} + b_{lm}r^{-(l+2)}]
               \frac{P_{lm}(\theta)}{\sin \theta} ime^{i m \phi}.
 $$
%\end{equation}
The
coefficients $a_{lm}$ and $b_{lm}$ are determined by imposing the
radial field at the surface from the Zeeman-Doppler maps and by
assuming that at some height $r=R_s$ above the surface (known as the {\em source surface}) the field becomes
radial and hence $B_\theta (R_s) = B_\phi(R_s) = 0$.  This second condition models
the effect of the plasma pressure in the corona pulling open field
lines to form a stellar wind. It imposes a relation between the
term in (\ref{psi}) that grows with radius and has coefficient $a_{lm}$ and the term that decays with radius and has coefficient $b_{lm}$ , since it implies that
\begin{equation}
\frac{a_{lm}}{b_{lm}} = -\frac{1}{R_s^{2l+1}}.
\end{equation}
Hence as the source surface radius decreases, the ratio $a_{lm}/b_{lm}$ increases and therefore the contribution to $\Psi$ of the part that grows with $r$ becomes relatively larger. As shown in Fig. (\ref{MeanB}), this means that at any given height, as the source surface is decreased, the average radial component of the field will increase while the non-radial contribution will decrease. Changing the source surface therefore modifies the field structure of the whole corona, such that, as shown in Fig. (\ref{fieldlines}), increasing the source surface lead to field lines that have a smaller radial component and are therefore rounder.

For each given surface map, therefore, the source surface is the one free parameter in determining the structure of the coronal magnetic field. Whatever value is chosen for this, however, the dominantly octupolar nature of the magnetic field close to the surface remains, as does the transition to a dipolar field at greater heights.  This transition can be seen in Fig. \ref{MeanB} where the radial field falls off with height as $r^{-5}$ (appropriate for an octupolar field) close to the star around 2$R_\star$, while a dependence closer to $r^{-3}$ (appropriate for a dipole) becomes apparent only at heights greater than around 5$R_\star$. Given the strengths of the dipolar and octupolar terms, we would expect the dipole to dominate at heights above 2$R_\star$.
 
\section{The accreting field}
Once we have determined the coronal structure of the magnetic field, we can divide the field lines into three types, depending on whether they are open, closed or possibly accreting. The only field lines that have the opportunity to accrete are those that pass through (or close to) the disk at a radius where the gravitational acceleration is directed towards the star. Field lines that never pass near to the disk region may either be open (i.e. wind-bearing) or closed (i.e. X-ray bright). Clearly, with our field extrapolation technique we cannot determine the fate of the field lines that pass through the disk. The nature of their interaction with the disk material will depend not only on the global disk properties, such as the mass accretion rate, but also on the small-scale physics of the plasma diffusivity and viscosity. An important factor is also the strength and geometry of the field lines that  intersect the disk. We show in Fig. (\ref{B_disk_nopress}) the range of field strengths of those field lines that pass through the disk. Models with a range of source surfaces are shown, some well within the Keplerian co-rotation radius and some beyond it. For the models where the source surface is inside the co-rotation radius, material accretes from the source surface along open field lines. In this case, the fall-off in field strength is just what would be expected from Fig. (\ref{MeanB}). For models where the source surface is beyond the co-rotation radius, however, we note that material can only accrete from within the  co-rotation radius. In this case, accretion takes place from a fixed radius along predominantly closed field lines whose field strength now varies much less with the value of the source surface. Field lines that penetrate the disk beyond the co-rotation radius are likely to be sheared open, adding to the flux of field lines that have been opened up by the pressure of the hot coronal gas. In this case our model would underestimate the flux of open field.

Considerable progress has already been made in studying the interaction of simulated stellar magnetic fields with accretion disks \citep{rekowski_3D_06,long_nondipole_07, matt_pudritz_torques_08,matt_pudritz_eqm_08}. Recent models of multipolar fields in particular show that accreting material may take the most energetically favourable path (i.e. closest to the equatorial plane) and so the form of the accretion funnels depends sensitively on the stellar field structure \citep{romanova_IAUS243_07}. The distortion of the field that will take place as it couples to the disk material is, however, beyond the scope of this paper and will be the subject of future work. Our present aim is to determine as a starting point which field lines are capable (by virtue of their location) of  connecting to the disk at all.

We therefore consider field lines that pass through the disc within
the co-rotation radius. We assume that the disc lies in the
equatorial plane and is of opening angle $10\degr$. The interaction of
the stellar magnetic field with the disc may lead to a warping of the
inner disc \citep{bouvier_PPV_06} in which case the disc may lie along
the magnetic equator, rather than the stellar rotational equator. In fact, 
we have found that allowing accretion to start from the magnetic equator
rather than the rotational equator makes little difference in this
case, and so in the absence of any evidence for a warped disc in V2129~Oph, we
consider the disc to lie in the plane defined by the stellar rotation
axis.  Along the field lines that pass through this plane inside the
co-rotation radius, we determine the density and velocity structure of
the accretion flow. 

%% fig4 
\begin{figure}
%\center{\includegraphics[scale=0.3]{denpro20.pdf}} 
%\center{\includegraphics[scale=0.3]{denpro40.pdf}} 
\center{\includegraphics[scale=0.3]{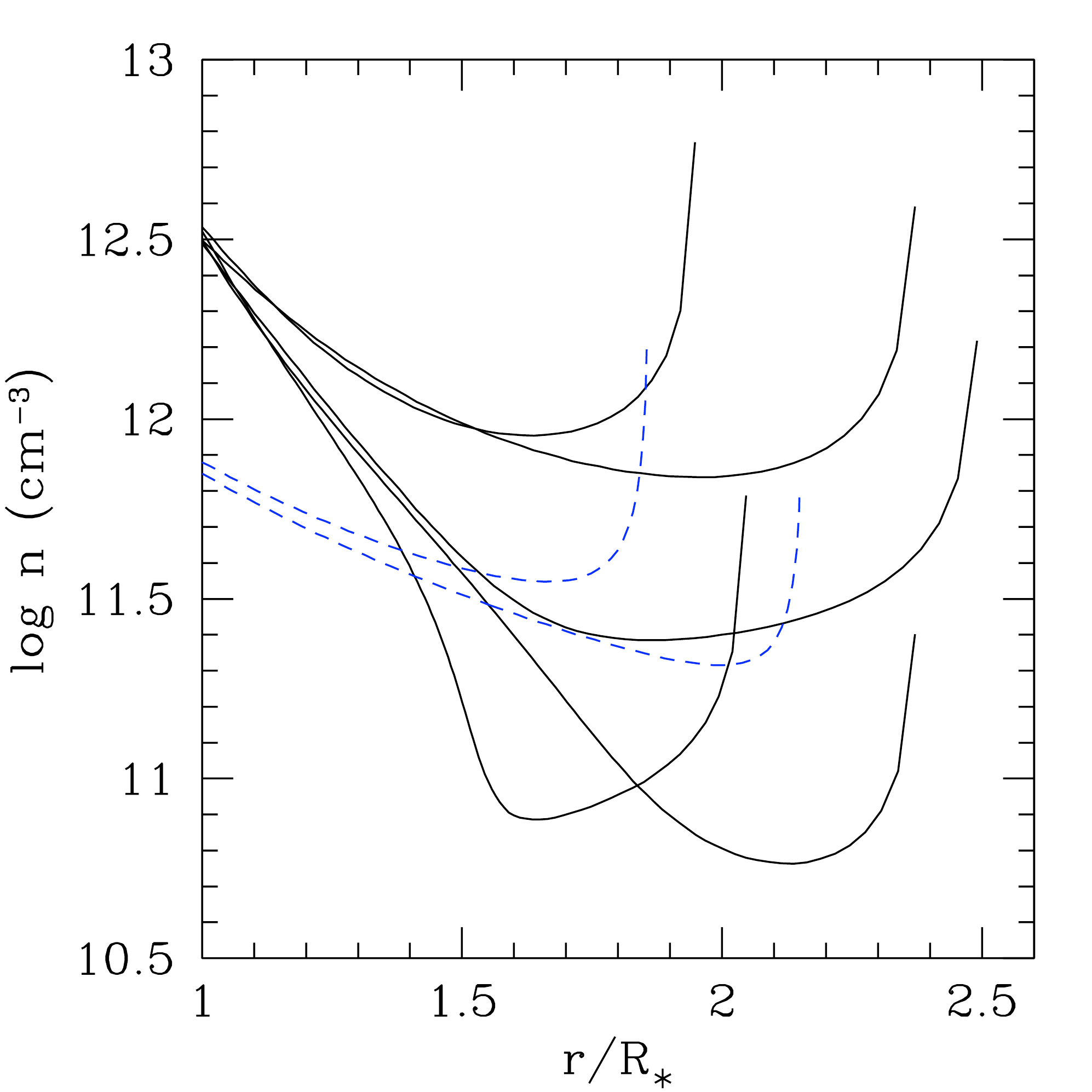}} 
\center{\includegraphics[scale=0.3]{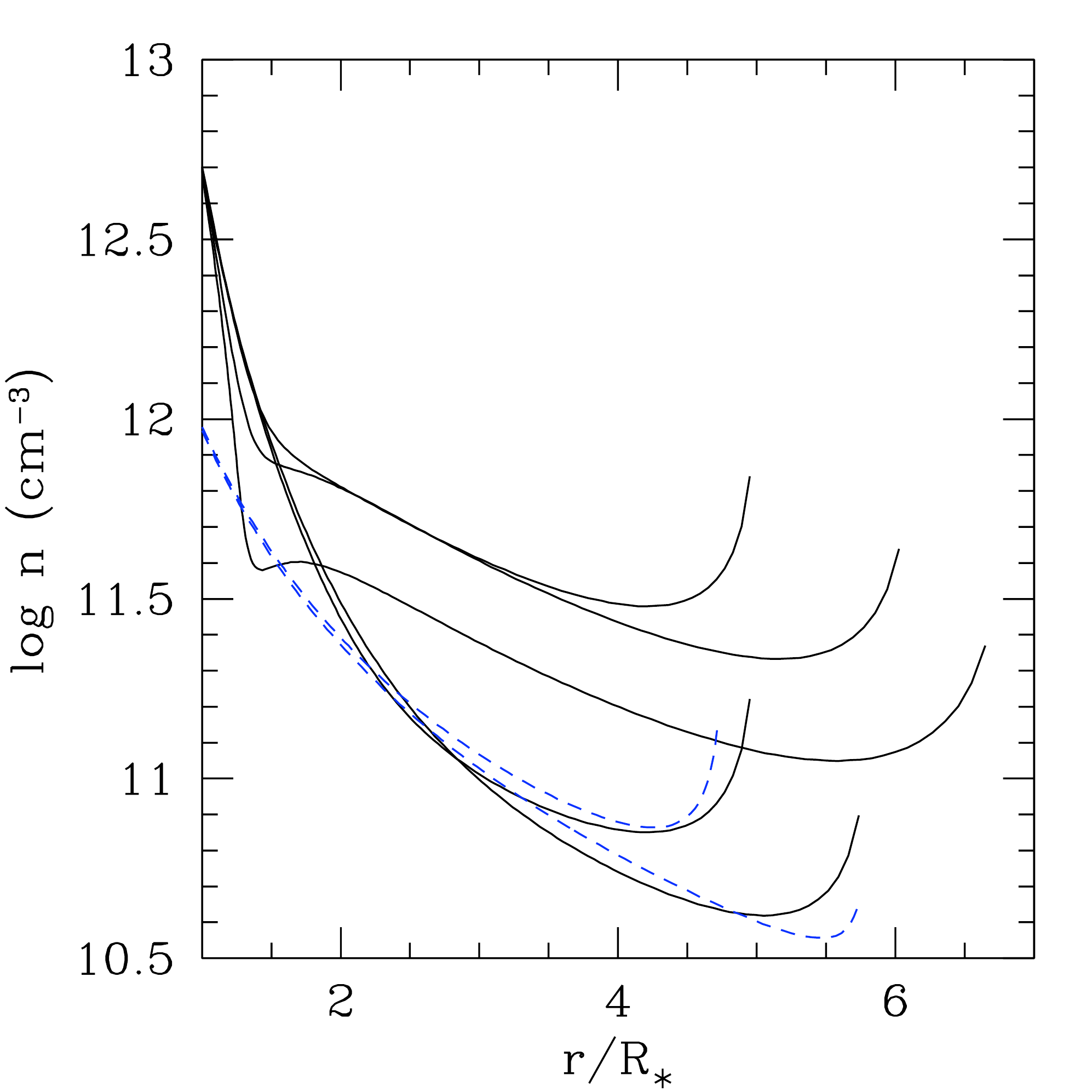}} 
\caption[]{Density profiles (solid lines) along example field lines. The source surface is at 2.5R$_\star$ (top) or 6.8R$_\star$ (bottom). The dashed lines show, for comparison,  density profiles along a dipole field. }
\label{denpro}
\end{figure}

%% fig5 
\begin{figure}
%\center{\includegraphics[scale=0.3]{vbar_error.pdf}} 
\center{\includegraphics[scale=0.3]{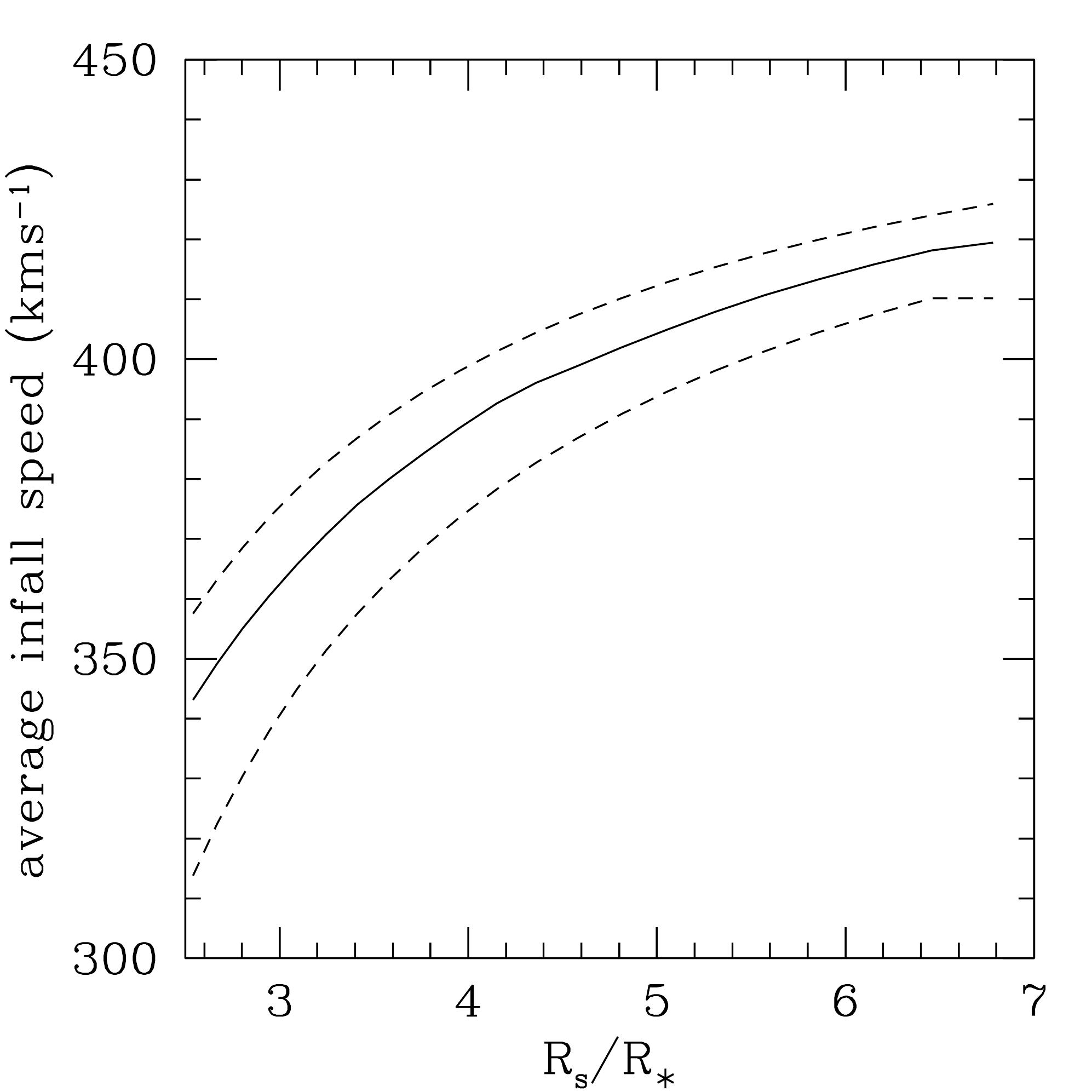}}
\caption[]{ Average infall speed (solid line) as a function of source surface radius. The dashed lines show the upper and lower limits of the range of infall speeds.}
\label{vbar_error}
\end{figure}

Following \citet{gregory_funnnels_07} we assume that the disc supplies mass at a
constant rate.  We use a spherical grid and assume that each grid cell
within the disc which is accreting supplies a mass accretion rate that
is proportional to its surface area.  For example, if an accreting grid
cell has a surface area that is $1\%$ of the total area of all accreting
grid cells, then this grid cell is assumed to carry $1\%$ of the total
mass accretion rate supplied by the disc.  Therefore, if grid cells which
constitute half of the total area of all accreting cells in the disc
carry material into a single hotspot, then half of the mass accretion
rate is carried from the disc to this hotspot.  In such a way the
accretion rate into each hotspot is different and depends on the structure
of the field connecting the star to the disc. Simulations suggest 
that both the mass and energy flux will be greatest at the centre of the
hotspot \citep{romanova_04}.

In order to calculate the variation in the density structure of the
accretion columns of V2129~Oph we assume that the field is
not significantly distorted by the flow and that the accreting gas flows
along the path of the field ($\bmath{v}$ is parallel to $\bmath{B}$) so
that
\begin{equation}
\frac{\rho v}{B} = const.
\label{mass}
\end{equation}
where $\rho$ is the mass density (see e.g. \citealt{mestel68}).  If an
individual accreting grid cell carries material onto an area of the
stellar surface of $A_{\ast}$ with velocity $v_{\ast}$ and density
$\rho_{\ast}$, then the mass accretion rate into $A_{\ast}$ is,
\begin{equation}
\dot{M} = \rho_{\ast} v_{\ast} A_{\ast}.
\label{mdot}
\end{equation}
It then follows from (\ref{mass}) and (\ref{mdot}) that the density
profile along the path of the field line may be written as,
\begin{equation}
\rho(r) = \frac{B(r)}{B(R_{\ast})}\frac{\dot{M}}{A_{\ast}v(r)}.
\end{equation}
The density profiles therefore do not depend on the absolute field
strength, but instead on how the field strength varies with height
above the star.  The density profiles are typically steeper than
those derived for accretion flows along dipolar field lines since
the strength of a higher order field drops faster with height above
the star (see Fig. \ref{denpro}).  Gas is
assumed to free-fall under gravity along the path of the field after
leaving the disc with a small initial velocity of
$10\,{\rm kms}^{-1}$ \citep*{muz01}.  The range of accretion velocities
with which flows arrive at the stellar surface is shown in Fig. \ref{vbar_error}. We assume that accretion
occurs over a range of radii, as discussed by \citet{gregory_funnnels_07}.  This
is equivalent to the approach taken previously by \citet*{har94},
\citet{muz01}, \citet{symington_tts_05}, \citet{aze06} and \citet{kur06} who
have demonstrated that such an assumption reproduces observed
spectral line profiles and variability.

In order to estimate the number density of gas within the accretion
columns we require the temperature distribution along the path of
the field lines.  The temperature variation within the accretion
columns of T Tauri stars is uncertain.  The most comprehensive model
is that of \citet{mar96} who considers both heating and cooling
processes.  Here we elect to take the simplest approach and assume an
isothermal accretion flow temperature of $7000\,{\rm K}$ - a
reasonable estimate for the temperature of accreting material
(see e.g. \citealt{har94}; \citealt{mar96}).  We obtain the number
density profiles from
\begin{equation}
n(r) = \rho(r)/\mu m_H,
\end{equation}
where $m_H$ is the mass of a hydrogen atom and $\mu$ the dimensionless
atomic weight.  Fig. \ref{denpro} show sthe
variation of the number density along the paths of a selection of
accreting field lines, with those obtained for dipolar field lines
shown for comparison.

The changing nature of the accreting field lines as the source surface is increased becomes very apparent when we consider the path taken by the accretion flows. Fig. (\ref{mdot_spot}) shows the fraction of the mass accretion rate that  impacts on the different regions of the stellar surface. When the source surface is very small, the accreting flow takes place along the complex smaller-scale field close to the stellar surface and impacts at a  range of locations. As the source surface becomes progressively larger, however, the accretion takes place along the simpler large-scale field until most of the accretion is divided equally between two accretion funnels, one in each hemisphere (see Fig. (\ref{acc_fieldlines_xray})). 
%% fig6
\begin{figure}
%\center{\includegraphics[scale=0.3]{mdot_spot.pdf}} 
\center{\includegraphics[scale=0.3]{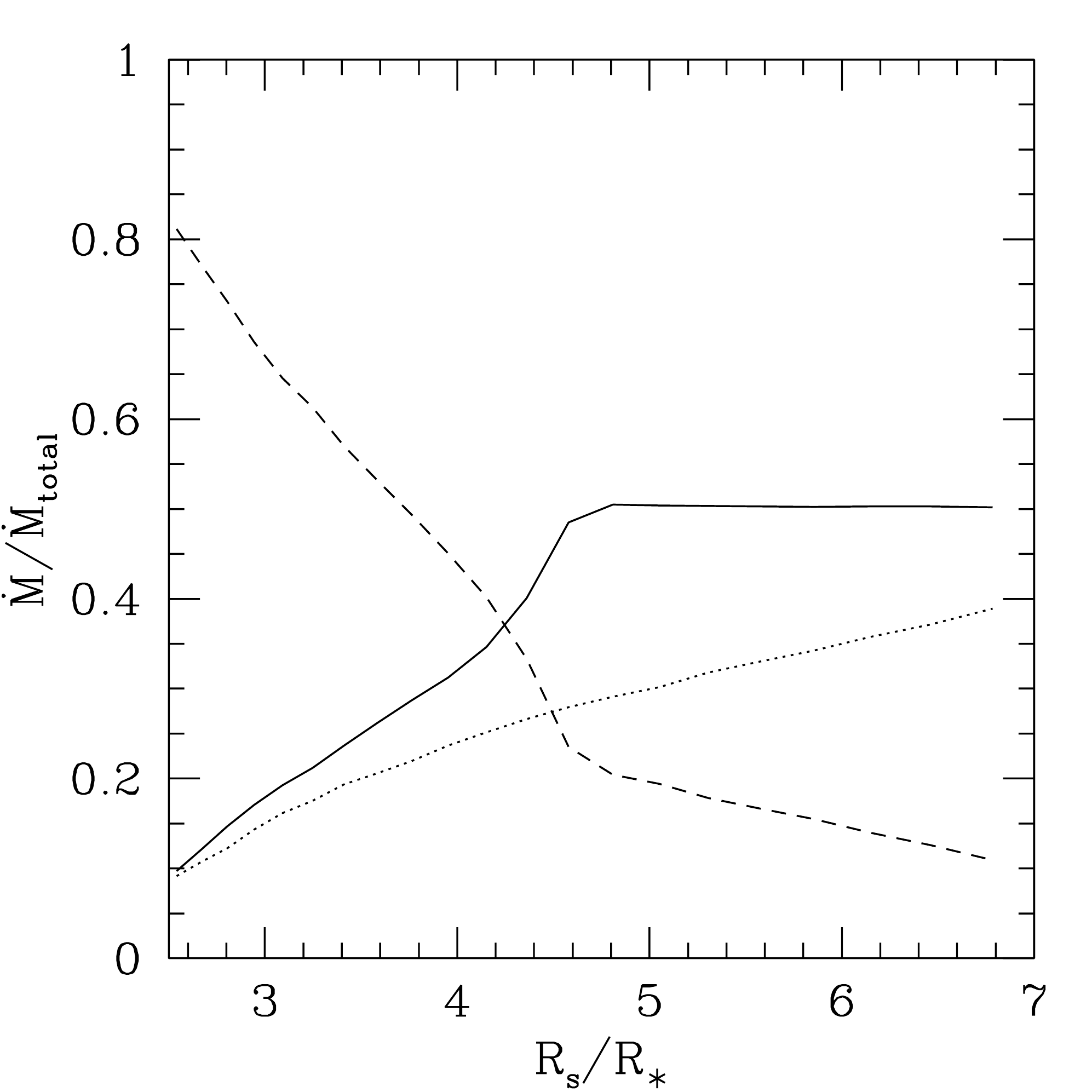}} 
\caption[]{The distribution of the total accretion rate into the
northern hemisphere hotspot (solid line), the southern hemisphere hotspot
(dotted line) and the various low-latitude hotspots (dashed line).}
\label{mdot_spot}
\end{figure}

%% fig7
\begin{figure*} 
\includegraphics[width=5.8cm]{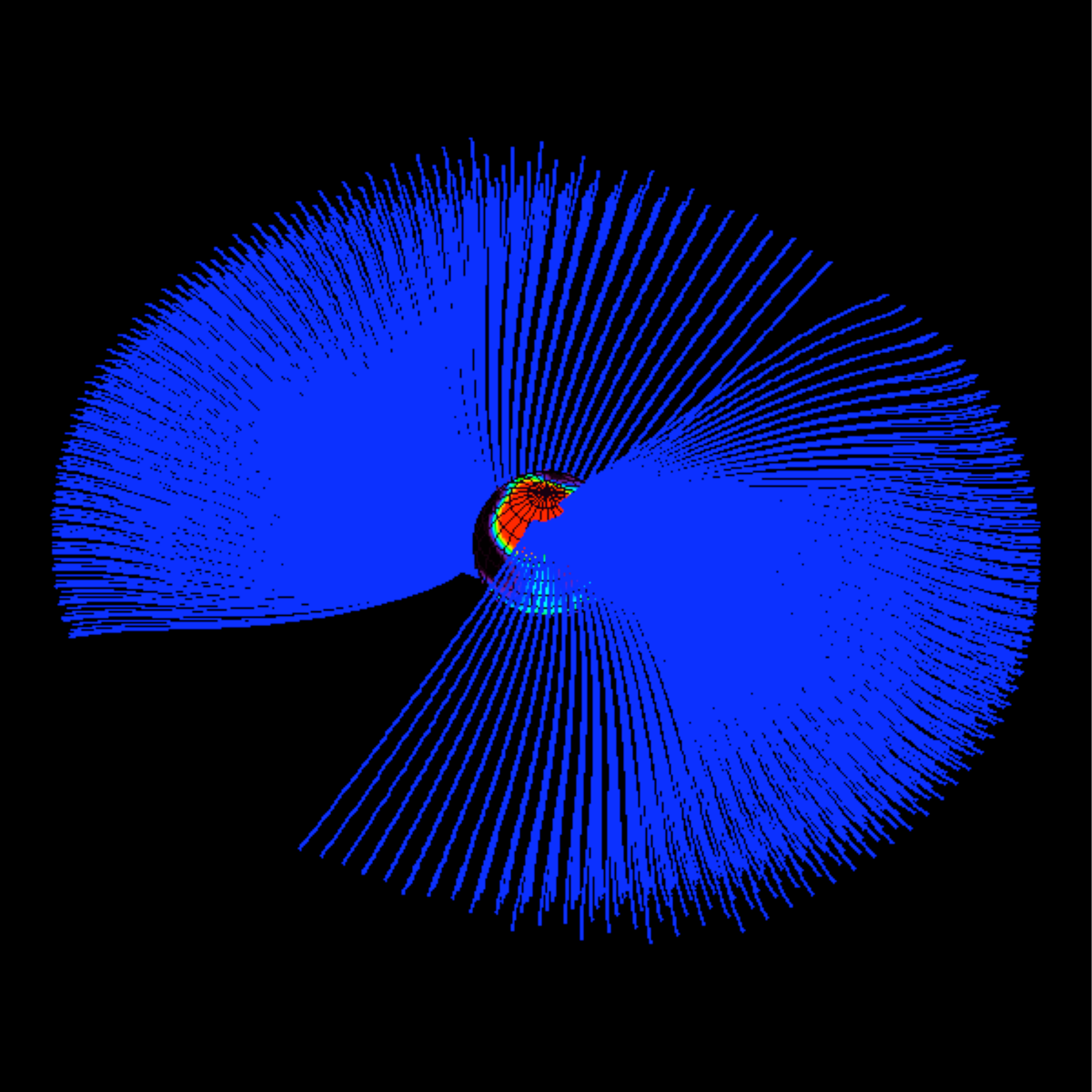}
\includegraphics[width=5.8cm]{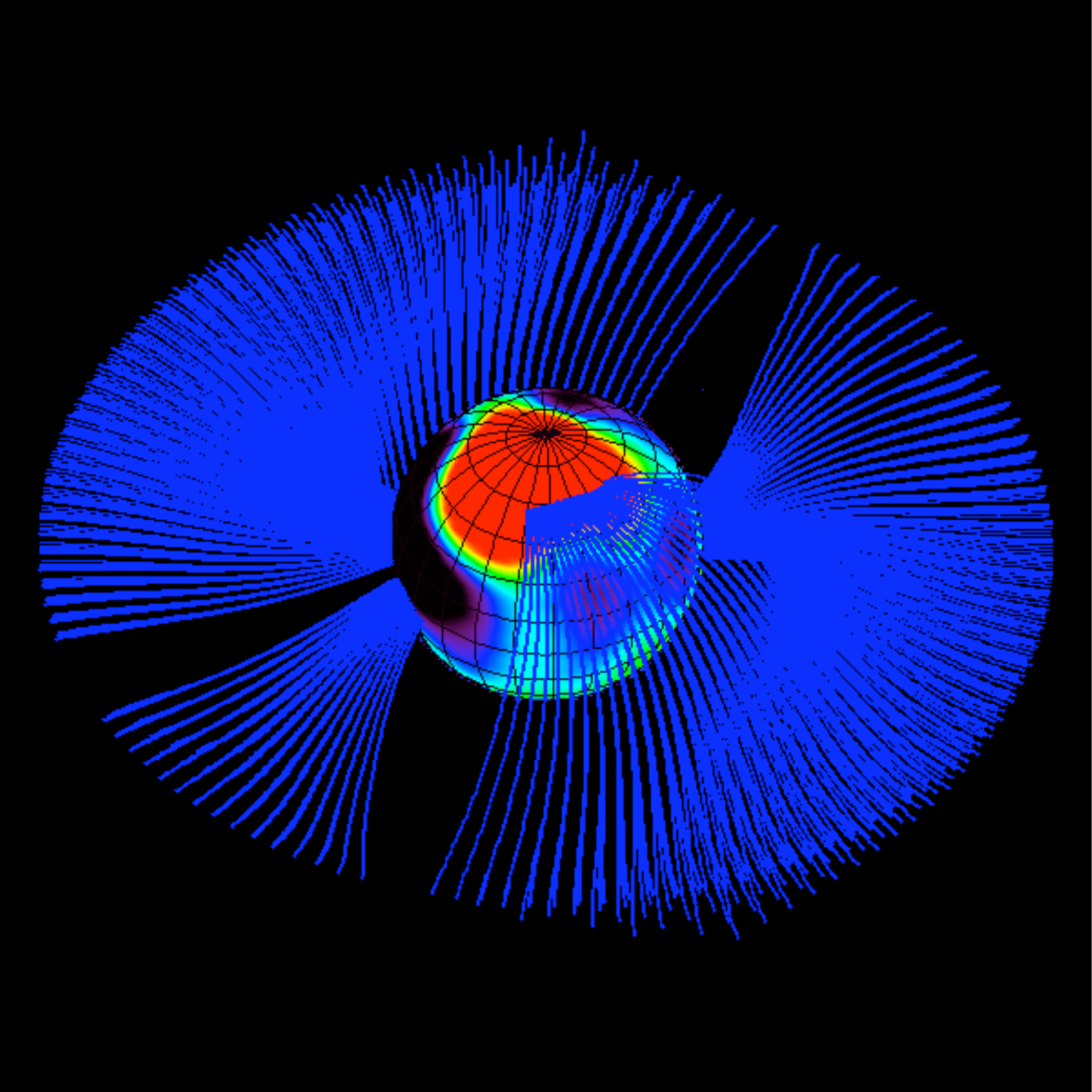} 
\includegraphics[width=5.8cm]{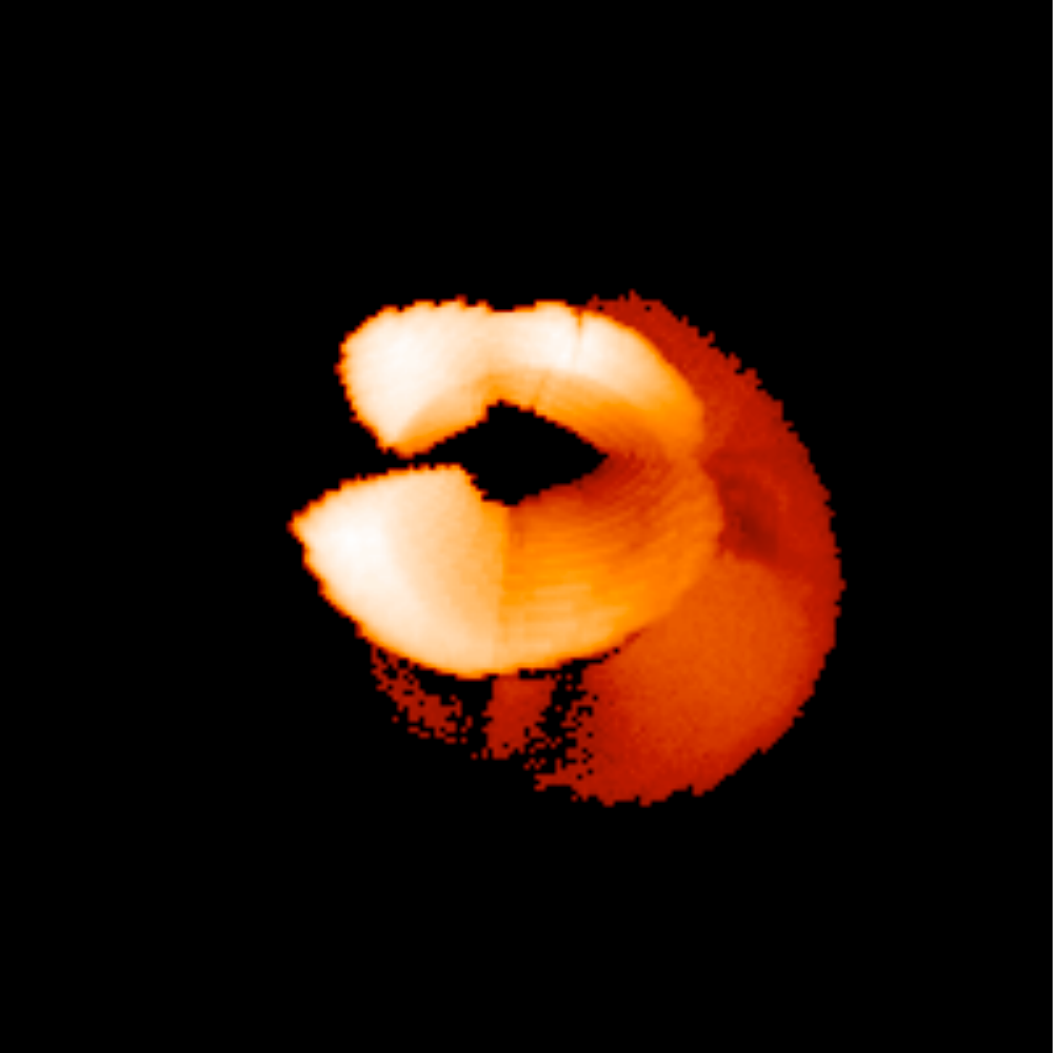}
\caption[]{Accreting field lines for V2129 Oph for a source surface at 6.8R$_\star$ (left) and at 3.2R$_\star$ (middle). The right-hand panel shows an X-ray image for a source surface at 3.2R$_\star$. In all cases the star is viewed at rotation phase 0.8 where the maximum in the accretion signatures is seen.}
\label{acc_fieldlines_xray}
\end{figure*}

\section{The X-ray bright field}
For the field lines that are closed and not interacting with the
disk we can calculate the pressure structure of the corona, assuming it to be
isothermal and in hydrostatic equilibrium.  Hence for a stellar
rotation rate $\omega$, the pressure at any point is $p=p_{0}e^{\frac{m}{kT}\int g_{s}ds}$ where $g_{s} =( {\bf g.B})/|{\bf B}|$ is the component of
gravity (allowing for rotation) along the field and
\begin{equation}
g(r,\theta) = \left( -GM_{\star}/r^{2} + 
                     \omega^{2}r\sin^{2} \theta,
		     \omega^{2}r\sin \theta \cos\theta 
             \right).    
\end{equation}
 At the loop footpoints we scale the plasma pressure $p_{0}$ to the
 magnetic pressure such that $p_{0}(\theta,\phi)=K
 B^{2}_{0}(\theta,\phi)$ where $K$ is a constant  that is the same on every field line.  By scaling $K$ up or down we can scale the overall level of the coronal gas pressure and hence the density and emission measure. The value of $K$ is a free parameter. Following \citet{jardine_tts_06} we choose the value of $K$ that gives the best fit when used to model the coronal emission of the stars in the COUP dataset. The plasma pressure
 within any volume element of the corona is then set to zero if the field
 line through that volume element is open, or if at any point along the field line the plasma pressure is greater than the magnetic pressure
 i.e. $\beta>1$.  From the pressure, we calculate the density
 assuming an ideal gas and determine the morphology of the optically
 thin X-ray emission by integrating along lines of sight through the
 corona.

 This technique has been used for some time now to extrapolate the
 coronal field of the Sun.  Comparisons of the resulting morphology of
 the corona with LASCO C1 images show good agreement for
 much of the corona, except near the polar hole boundaries where the
 potential field approximation is likely to break down at the
 interface between open and closed field regions \citep{wang97}.  \citet{wang97}
 found the best agreement with observations was obtained with a
 scaling of $p_{0} \propto B^{0.9}_{0}$.  The question of the optimal
 scaling is one that is often addressed in the context of the heating
 of the solar corona (see \citet{aschwanden2001} and references
 therein) and indeed recent work suggests that scaling laws developed
 on the basis of detailed solar observations may be extrapolated to
 more rapidly rotating stars \citep{schrijver01}.  The use of field
 extrapolation based on surface magnetograms obtained using
 Zeeman-Doppler imaging has been developed extensively for 
 rapidly-rotating main-sequence stars \citep{jardine02structure,jardine02xray,mcivor06xray}. Comparison of the predicted coronal X-ray emission with Chandra line shifts and emission measures confirms the usefulness of this technique \citep{hussain_letgII_07}.
 
%% fig8 
\begin{figure}
%\center{\includegraphics[scale=0.3]{emrange_accn.pdf}} 
%\center{\includegraphics[scale=0.3]{emrange_noaccn.pdf}} 
\center{\includegraphics[scale=0.3]{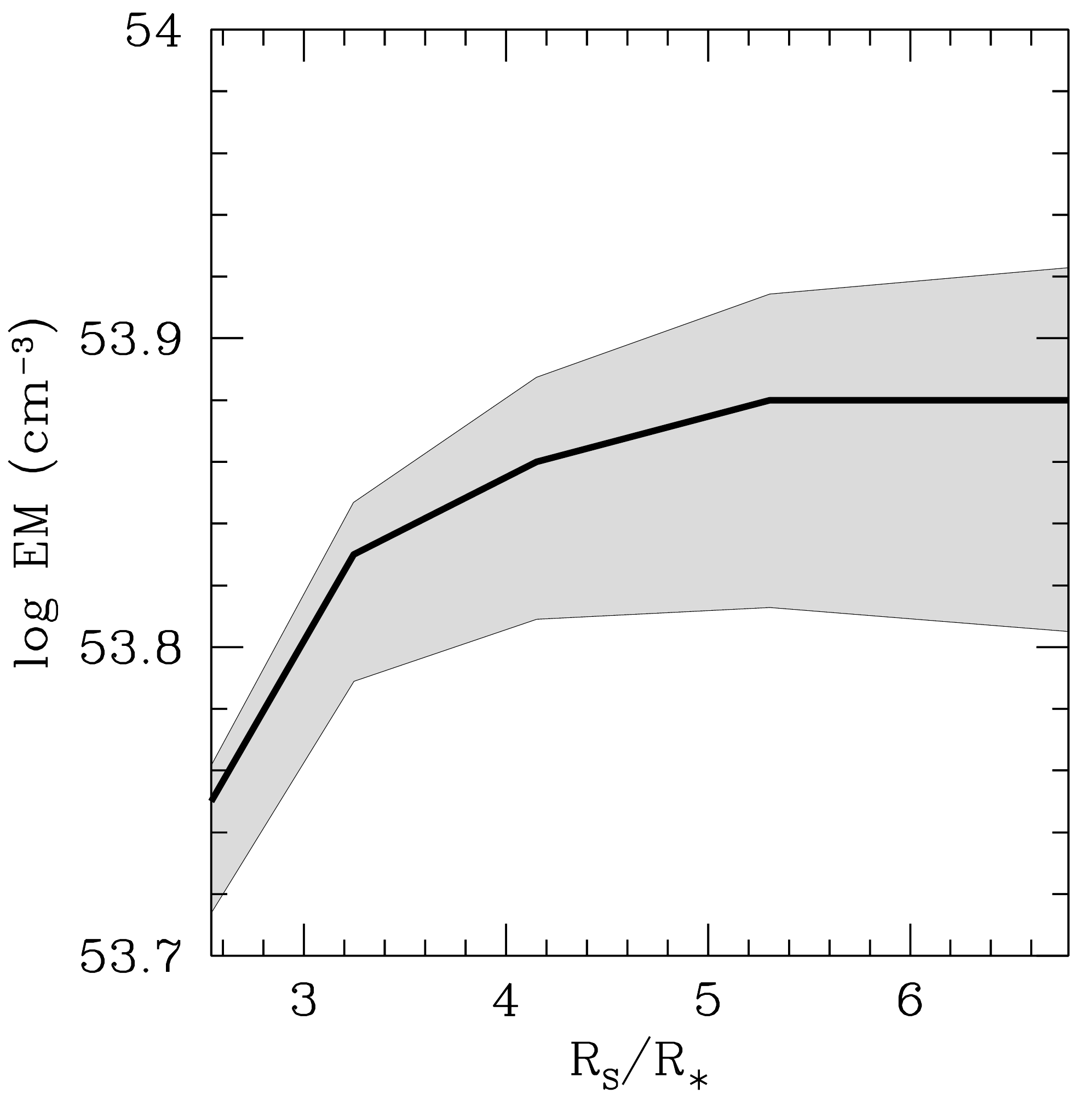}} 
\center{\includegraphics[scale=0.3]{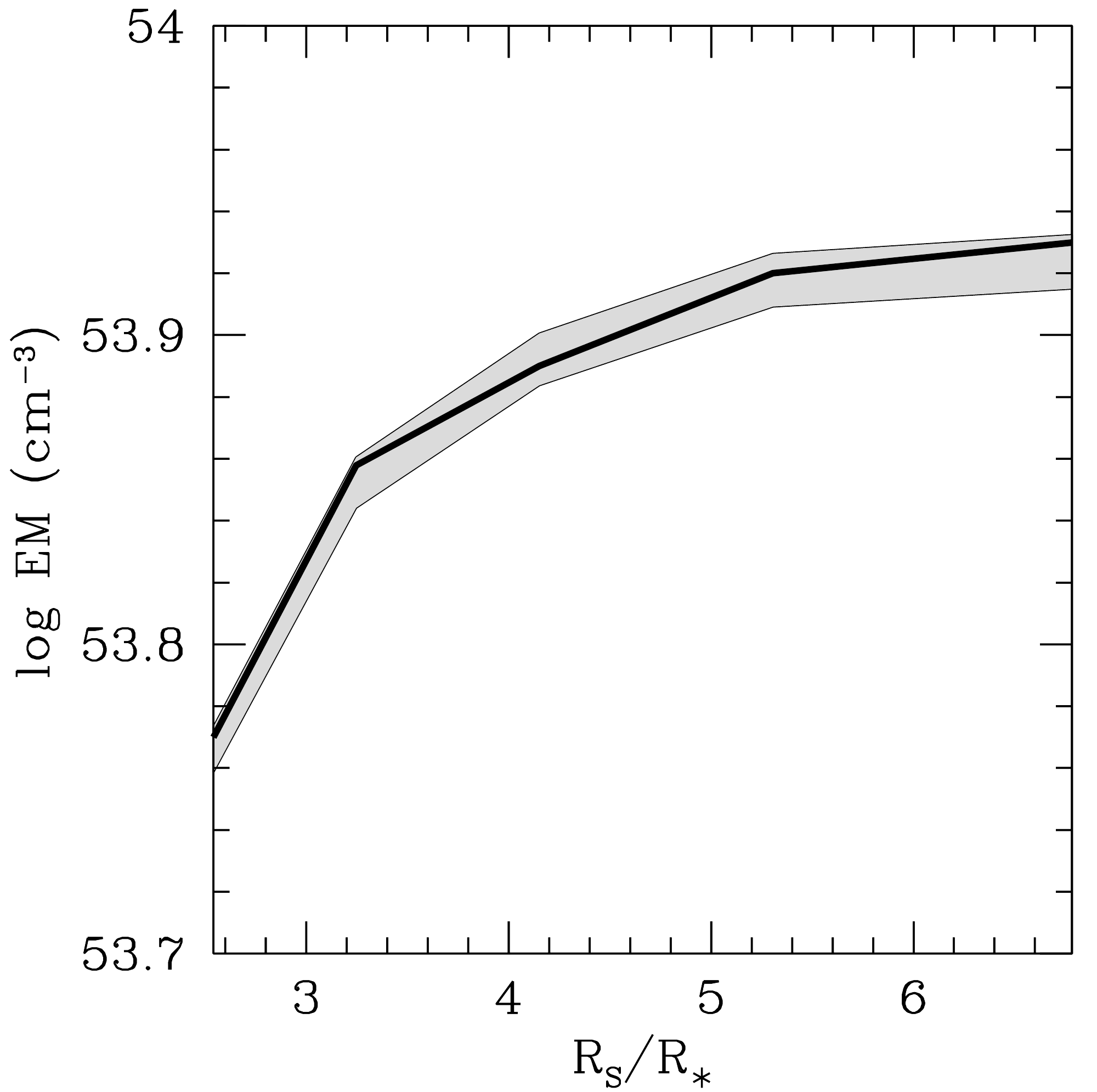}} 
\caption[]{Calculated emission measure for various source surfaces when scattering by accretion funnels is taken into account (top) or neglected (bottom). The shading demonstrates the range of emission measures over one stellar rotation.  }
\label{emrange}
\end{figure}

One significant difference, however, between these main-sequence stars and T Tauri stars is the presence of active accretion. The disk may truncate the corona, forcing field lines that might otherwise have been closed (and hence X-ray bright) to be open or accreting, and hence X-ray dark \citep{jardine_tts_06}. In addition, the accretion funnels themselves may scatter some of the X-ray photons from the corona, causing not only a drop in the magnitude of the X-ray luminosity, but also an increase in its rotational modulation. Using realistically-complex magnetic field structures, \citet{gregory_rotmod_06,gregory_funnnels_07} have shown that for the actively-accreting T Tauris in the COUP sample, the observed drop in $L_x$ (and its rotational modulation) could be explained by the presence of dense accretion funnels. 

In order to test this hypothesis with the derived magnetic structure for V2129 Oph, we have calculated the density profiles along the accreting field lines and determined the X-ray emission measure as a function of rotational phase for a range of models with different source surfaces. We chose a coronal temperature of 3$\times$10$^7$K which is typical for stars in the COUP database. A lower temperature would have given a lower emission measure, but the behaviour as the source surface is changed would have been qualitatively unchanged. The values of the emission measure, while high for most solar-mass T Tauri stars, are typical for a star of this mass. The average value for the case of a source surface at 6.8R$_\star$ is $7.6\times10^{53}$~\pcc\ which translates to an X-ray luminosity of  $6.7\times10^{30}$~\eps. This  compares well with the published value of  $2.5\times10^{30}$~\eps\  which was derived from ROSAT observations which were sensitive to rather cooler plasma \citep{casanova_95}.

The effect of the accretion funnels can just be seen in the X-ray image in Fig. (\ref{acc_fieldlines_xray}) which is viewed at rotation phase 0.8 which coincides with the inferred location of the accretion hotspots. The effect on the X-ray emission is rather small because of the relatively low density of the accretion funnels, 
and the fact that the magnetic geometry of V2129~Oph is such that there is a ring of X-ray bright field 
at high latitude which is barely obscured by the channelled funnel of accreting gas.
Fig. (\ref{denpro}) shows the variation in the density for some sample field lines, for a model with a source surface at 6.8R$_\star$ and one at 2.5R$_\star$. In each case, the corresponding density profile for dipolar field lines is also shown. The larger the source surface becomes, the more dipolar do the density profiles become. In both cases, however, the density close to the stellar surface is similar and is of a magnitude such that scattering from the accretion funnels can reduce the X-ray luminosity, although the predicted rotational modulation for this moderate accretor is only of order 10$\%$ at most. Fig. (\ref{emrange}) shows the corresponding X-ray emission measure for this set of models when the accretion funnels are present and also when they are neglected. In both cases, increasing the source surface initially increases the emission measure as the volume of the closed corona increases, but once the source surface extends beyond the co-rotation radius, the maximum extent of the closed corona is limited to the co-rotation radius and so the emission measure becomes independent of the source surface. 

The degree of rotational modulation of the X-ray emission measure also varies with the value of the source surface. As  Fig. (\ref{acc_fieldlines_xray}) shows, however, the octupolar structure of the field  results in a ring of emission at high latitudes. Due to the inclination of the star ($i=45^\circ$) this ring is almost always in view as the star rotates, and hence, as can be seen from Fig. (\ref{emphase2}), the rotational modulation is small. When the source surface is placed at larger radii (in this case close to the co-rotation radius) both the magnitude and the rotational modulation of the X-ray emission are greater. In the case where scattering due to the accretion funnel is considered, the location of the minimum also varies slightly, being closer to phase 0.8 when the source surface is larger. The difference is simply due to the alteration in the shape of the field lines as the source surface is changed. In this case, the field lines that are accreting connect to the star at slightly higher latitudes and later phases when the source surface is moved out.

%% fig9 
\begin{figure}
%\center{\includegraphics[scale=0.3]{emphase2.pdf}} 
\center{\includegraphics[scale=0.3]{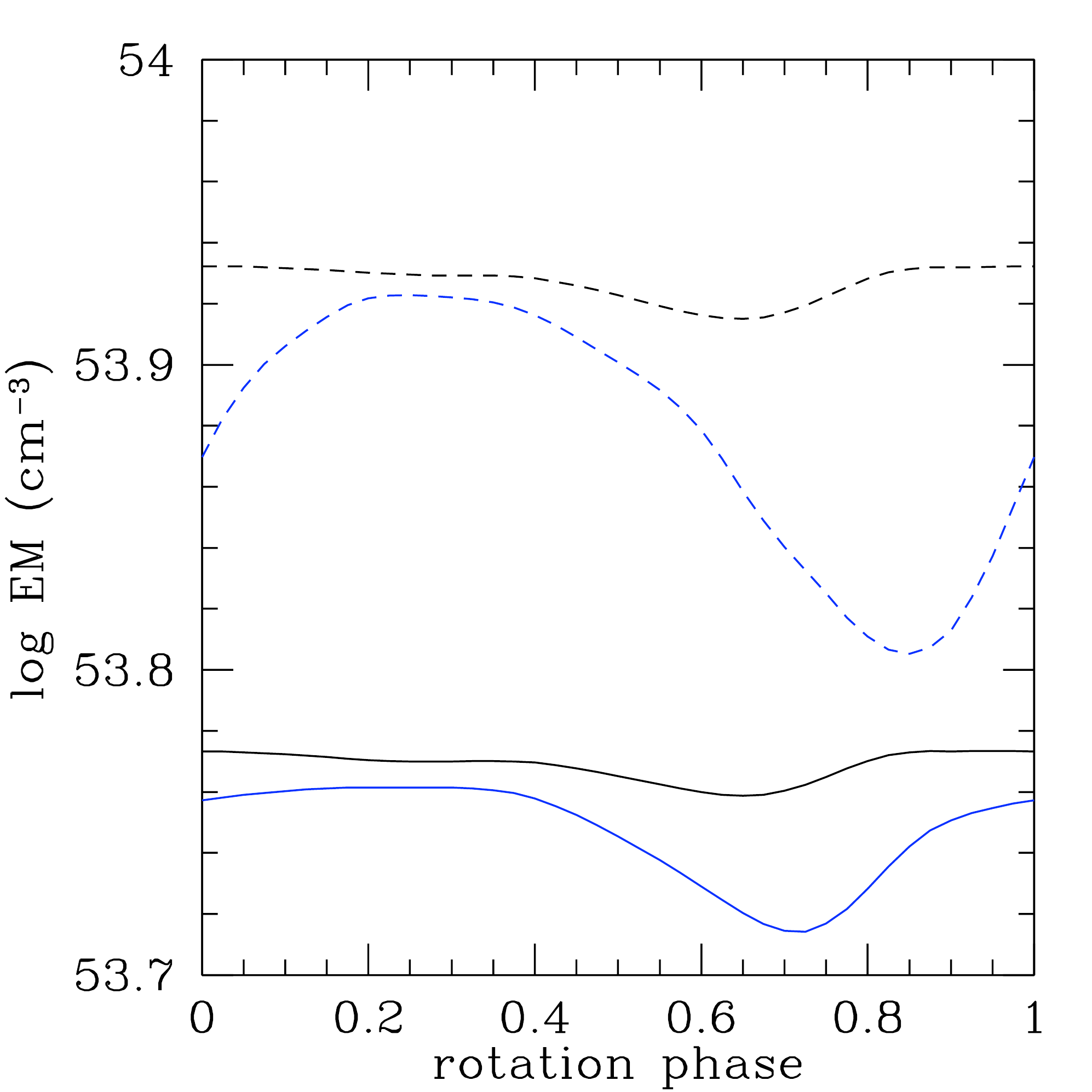}} 
\caption[]{Rotational modulation of the emission measure. The dotted lines are for a source surface at 6.8$R_\star$ and the solid lines are for a source surface at 2.5$R_\star$.  Black lines show the case where scattering by accretion funnels is neglected, while blue lines show the case where scattering by accretion funnels is considered.  }
\label{emphase2}
\end{figure}

\section{results}

From the surface magnetogram of V2129~Oph, we have extrapolated the coronal magnetic field, and from this calculated the location, density and velocity of the accretion flows and also the coronal density distribution and X-ray emission measure. In doing this we have two free parameters. One is the constant of proportionality $K$ that determines the plasma pressure at the base of the corona, and hence the overall magnitude of the X-ray emission measure. In the absence of observations of V2129 Oph that could determine these parameters, we choose to select the value of $K$ that gives the best fit to the range of emission measure of the stars in the COUP sample.  The other free parameter of our model is the value of the source surface where the outward pressure of the coronal gas is assumed to have opened up all of the magnetic field lines. We have taken a range of values for the source surface, from one well inside the co-rotation radius (which ensures that all field lines at co-rotation are open) to one well beyond the co-rotation radius (which ensures that there is a mixture of open and closed field lines at co-rotation). 

The principal difference in the accretion process as we increase the value of the source surface is that accretion takes place along progressively simpler field structures and impacts on progressively fewer sites at the stellar surface. We find that in order to have the majority of the accretion impacting on a single high-latitude site in each hemisphere, we need to place the source surface beyond approximately 7R$_\star$, close to the Keplerian co-rotation radius at 6.8R$_\star$. At this distance from the star, the magnetic field lines that intersect the disk have a strength of only a few Gauss and are certainly weak enough to be overwhelmed by the ram pressure of the accreting gas.

As the source surface is increased and the extent of the closed-field corona increases, we find that the magnitude of the predicted X-ray emission measure increases, and its rotational modulation also increases. Rotational modulation can be caused either by self-eclipse of X-ray bright regions, or by scattering from the accretion funnels themselves. For V2129 Oph, this predicted rotational modulation is small (only around 10$\%$ at most) partly because the X-ray bright regions remain in view as the star rotates, and partly because it is only a moderate accretor and so the density in the accretion funnels is relatively small.

\section{Conclusions}
%% Bibliography

V2129~Oph is a young star in which a small radiative core appears to have already formed. For this star, the dynamo produces a field that is predominantly antisymmetric, with a complex surface distribution of polarities in which the strongest mode is that of a 1.2kG octupole. A weaker dipole component is also present, with polar strength of 0.35kG . 
 By extrapolating this surface field into the corona, we have been able to determine the form of the field at a range of distances from the surface. The observation that accretion takes place mainly at one range of longitudes, into a region of positive polarity at the surface, suggest that at the field at the inner edge of the disk is much simpler than the surface field. This constrains the rate at which the field complexity decays with height and sets the limit of the closed-field corona at a height close to the co-rotation radius of 6.8R$_\star$ which is consistent with traditional disk-locking models for star-disk interaction \citep{konigl91,cameron_campbell_93}.  This is also consistent with the predictions of \citet{jardine_tts_06} that for a T Tauri star of this mass, the closed-field corona should extend out to approximately the co-rotation radius. The fact that such a relatively weak dipolar component is able to trigger accretion as far out as the co-rotation radius remains a challenge for accretion models which have traditionally considered purely dipolar fields. Progress  is being made, however, in modelling more complex field topologies \citep{long_nondipole_07}. 

Thus, although  Fig. \ref{acc_fieldlines_xray} shows that the X-ray emission is concentrated within  a height of ~1R$_\star$ above the surface, the closed-field corona extends out much further. The degree of field complexity that might be inferred by tracing the X-ray bright loops is much greater than that seen in the accreting loops. This has parallels in the solar case, where the X-ray corona is compact and very highly structured, while the K-corona, which is also composed of gas at 10$^6$K trapped on closed field lines, is much smoother and extends out to typically 2.5R$_\odot$. The solar K-corona is not detected in X-rays because the density of the gas at these distance is too low. In a similar manner, V2129 Oph also appears to produce a large-scale, simple corona which is not apparent in the X-ray images, but which is in some sense ``illuminated'' by the accretion flows along it, without which it may have passed undetected.

More observations of T Tauri stars of different masses and evolutionary states are needed to determine if this field structure is particular to young stars with radiative cores, or if it is also seen on fully-convective stars which may produce magnetic fields through a different type of dynamo. Simultaneous phase resolved X-ray and spectropolarimetric observations are also needed to determine if X-ray light curve eclipses are indeed visible in accreting systems and if they occur at the phases predicted on the basis of the derived magnetic field structures.

%\bibliographystyle{mn2e}
%\bibliography{iau_journals,mmjpapers}

%\bibliography{v2129oph}

%\end{document}
%------------------------------------------------------

%------------------------------------------------------

\end{document}